\documentclass[slac_one,floatfix]{revtex4}
\usepackage{graphicx}
\usepackage{fancyhdr}
\usepackage{lscape}
\begin{document}

\title{{\small{Report to the ILC World-wide Study}}\\ 
\vspace{12pt}
Physics Benchmarks for the ILC Detectors} 

%

\author{M. Battaglia}
\affiliation{Dept. of Physics, University of California  and LBNL,
Berkeley, CA 94720 USA}
\author{T. Barklow, M. E. Peskin}
\affiliation{SLAC, Stanford CA 94309 USA}
\author{Y. Okada}
\affiliation{KEK, Tsukuba-shi, Ibaraki-ken 305-0801  JAPAN}
\author{S. Yamashita}
\affiliation{International Center for
Elementary Particle Physics, 
University of Tokyo,
Tokyo 113-0033  JAPAN}
\author{P. Zerwas}
\affiliation{DESY, Hamburg  D-22603 GERMANY}

\begin{abstract}
This note presents a list of physics processes for benchmarking the performance 
of proposed ILC detectors.  This list gives broad coverage of the required physics 
capabilities of the ILC experiments and suggests target accuracies to be achieved.
A reduced list of reactions, which capture within a very economical set the 
main challenges put by the ILC physics program, is suggested for the early stage of 
benchmarking of the detector concepts.

\end{abstract}

\maketitle

\thispagestyle{fancy}


\section{INTRODUCTION} 

Now that the RF technology of the ILC has been chosen, concepts are
taking shape for the design, components, and optimization of the 
detectors~\cite{ldc,sid,gld}.   At the same time, a program of R\&D on 
sensors and sub-detector components is being carried 
out worldwide.   It is important to both studies to justify the design
R\&D goals by the required physics capabilities of the detector.
In time, it will also be important to compare the relative integral
performance of the different concepts.  These evaluations should be 
based on a standard set of 
reference physics processes.   Even today, at the early stages 
of the detector design process, it is valuable to review the accuracy targets
that must eventually be met for the major reactions of the linear collider
physics program and to survey the full set of capabilities that a linear
collider detector must have.  In this note, we would like to propose a 
set of reference reactions that addresses these goals.  We hope that this
list of reactions will be a useful guidepost in the R\&D studies.  We also
hope that, by providing focus on specific problems, 
 it will promote more realistic physics simulations that take
detailed account of the detector response and the machine-detector interface.
This will be needed for the detector designs, and it will also help in 
refining our assessment of the ILC physics potential.

A formal set of benchmark processes should fulfill some basic requirements.
First, it should be designed so that the central physics scenarios are 
broadly covered.  It should contain the most important reactions that give
the physics justification of the ILC and illustrate how the ILC complements
and extends the results expected from the LHC.  Second, the benchmark 
processes should be robust.  A benchmark should not be tailored to a singular
physics problem but rather should address issues common to a variety of physics analyses.  
That is a benchmark should ensure that the detector criteria that it tests would be 
appropriate also for physics scenarios that 
have not been considered.  Third, the effect of the performance of individual 
detector components on the physics results should be manifest. 
It is important that the benchmarks have well-defined target goals for 
measurement accuracy that are motivated quantitatively by the requirements
of the ILC physics program and the anticipated potential of the LHC and of 
astrophysics experiments.

These requirements can be addressed by examples chosen from three classes of
processes: i) studies of the Higgs boson or Higgs sector;
ii) studies of Supersymmetry,  and iii) precision measurements in the
Standard Model with indirect sensitivity to New Physics. Examples in the
third class also give precise goals for measurements of beam energy and collider 
luminosity, measurements that will also be crucial in the precision study of 
new and exotic particles.

Several sets of benchmarks have 
been proposed over the years of the linear collider physics study.
However, we feel that, as we begin a new stage in the definition of the
ILC and its detectors, it is useful to comprehensively review the 
menu of benchmark processes.   In this note we  define  new
sets of physics reactions that we propose be used in  validating the 
detector design 
and optimising its performances. This note is organised as follows. We first 
discuss a broad set of key physics reactions that met the goals just 
discussed for broad coverage of the ILC program, and that
have well defined target performances. We then review the physics-detector 
matrix linking the performance of the various detector subsystems to the 
physics reactions. Finally, as a starting point for detailed studies of 
detector design, we propose a set of ILC measurements that offer an adequate 
sampling of the major challenges for the ILC detectors within as small a list 
of processes as possible.

\section{PHYSICS REACTIONS}

In this section,  we review physics processes of interest in defining detector
performance.  For each quantity to be measured, we present a target 
accuracy.  These accuracies are based on quantitatively
well-defined goals of the ILC, whose limits are set either by 
theoretical uncertainties
(as in the case of the Higgs branching fractions) or by 
matching the accuracy of 
other measurements (as in the case of the cosmology-motivated 
SUSY scenarios).  Each goal is to be obtained by running at one of
three fixed centre of mass energies ---350~GeV, 500~GeV, or 1~TeV---with 
the event samples
expected from the ILC design---0.5~ab$^{-1}$, 0.5~ab$^{-1}$, 
1.0~ab$^{-1}$, respectively.  In most cases, the reactions listed have
been studied using parametric simulation and the required target accuracies
have been shown to be achievable at that level.

\subsection{Higgs Boson and Higgs Sector}

The study of the Higgs sector offers a set of reactions where physics 
results need to be extracted with clearly defined accuracy.  These 
reactions involve a wide variety of final states, ranging from two leptons 
to eight jets.   The particular case of a Standard Model Higgs boson $h^0$ 
with mass $M_h = 120$~GeV offers significant branching fractions to a large 
variety of different decay channels.  Through this case, we can sample the 
main processes that will define the profile of a light Higgs boson. 

\subsection{Supersymmetry}

Supersymmetry offers an enormous parameter space in which many different
physics processes appear at varying levels of importance.   It is often 
not obvious what accuracy is required for measurements of supersymmetry
particle masses and properties.  However, there are two circumstances 
in which the requirement are very clear.  First, when some parameters of
the superparticle spectrum can be measured at LHC, one should insist
the new information from ILC signficantly improves the accuracy of this
analysis.  Second, when the supersymmetry parameters are such as to 
produce a candidate for the cosmic dark matter, one should insist that the 
linear collider experiments should allow one to compute the dark matter
density to an accuracy that would provide a significant comparison
with the determination to a few percent accuracy from measurements of the
cosmic microwave background.  Both arguments turn out to require high-precision
mass measurements, at the level of 1\%--0.1\% accuracy, for several 
supersymmetry particle masses or mass differences.  The various scenarios
that we have included
for supersymmetric dark matter emphasize different final states that range
from dileptons plus missing energy to complex
multi-jet events.   In addition, we have included three special models
with late-decaying particles or exceptionally small mass degeneracy.
Parameters of the mSUGRA benchmark points are given in Table~\ref{SUSYtable}, 
those of the alternative SUSY breaking mechanisms are given in the footnotes
\{20\} and \{21\}.

\begin{table} 
 \caption{\label{SUSYtable}  Model parameters and particle masses for benchmark points
 taken as points in the mSUGRA model.  The spectrum for these points have been computed 
 using ISAJET 7.69~\cite{Paige:2003mg} as a reference code.}
\begin{tabular}{|c|l|c|c|c|c|c|c|c|c|c|c|c|c|c|c|}
\hline 
Point & Ref. & $m_0$ & $m_{1/2}$ & $\tan \beta$ & $A$ & $\mu$ & $M_{top}$ &
$M_{\tilde \chi^0_1}$ & $M_{\tilde \tau_1}$ & 
$M_{\tilde \chi^0_2}$ & $M_{\tilde \chi^0_3}$ & $M_{\tilde e_R}$ & $M_A$ & 
$M_{\tilde \chi^+_1}$ & $M_{\tilde \chi^+_2}$ \\
  &  & GeV & GeV & GeV &  & GeV & GeV & GeV &  GeV & GeV & 
GeV  &  GeV & GeV & GeV & GeV \\
\hline \hline
1 & {\tiny SPS1a'}~\cite{Allanach:2002nj,lccosmo} & ~70 & 250 & 10 & -300 & 
389 & 175 & ~96.1 & 109.2 & 185 & 393 & 124 & ~421 & 185 & 408 \\
2 & {\tiny LCC2}~\cite{lccosmo}   & 3280 & 300 & 10 & 0 & 178 & 175 & 
107.7 & 3251 & 166 & 190 & 3270 & 3242 & 159 & 287\\
3 & {\tiny D'}~\cite{Battaglia:2003ab}  & 110 & 525 & 10 & 0 & 654 & 175 &
211.3 & 220.8 & 408 & 658 & 228 & 744 & 409 & 671 \\
4 & {\tiny LCC4}~\cite{lccosmo}  & ~380 & 420 & 53 & 535 & 0 &
178 & 169.1 & 195 & 327 & 540 & 412 & ~419 & 328 & 553 \\ 
5 & $\alpha$~\cite{DeRoeck:2005bw} & 206 & 293 & 10 & 0 & 375 & 178 & 113 & 213 & 
215 & 380 & 216 & 265 & 215 & 399 \\ \hline \hline
6 & $\epsilon$~\cite{DeRoeck:2005bw} & 20 & 440 & 15 & -25 & 569 & 178 & 175 & 
153 & 339 & 574 & 171 & 622 & 340 & 587 \\ \hline
\end{tabular}
\end{table}

\subsection{Precision Tests of the Standard Model}

Some models of new physics that do not involve supersymmetry
nevertheless lead to signatures similar to those of supersymmetric
models.  But other new physics scenarios lead to high-mass 
resonances that affect the standard process of $e^+e^-$ annihilation
indirectly.  For example, models in which the Higgs bosons are heavier than the
precision electroweak limit typically contain new vector resonances.

The ILC has very powerful capabilities
to analyze such models through precision measurements on the 
affected processes.  The sensitivity of these measurements to 
heavy resonances typically exceeds the sensitivity of the LHC to 
observe the resonance directly.  Further, if a heavy resonance is observed
at the LHC, precision measurement of the anomalies it produces in Standard 
Model reactions will give us detailed information about its couplings.
Thus, we have included precision measurements of 
all of the major Standard Model processes as elements of our list of 
benchmark reactions.  We have also included two scenarios that illustrate 
different manifestations of new physics, the ADD scenario of large extra 
dimensions leading to single-photon missing energy events, and a scenario
in which a graviton resonance in the Randall-Sundrum scenario 
is directly accessible at 1~TeV.

\begin{turnpage}
\begin{table}[h!] 
\begin{center}
\caption{\label{BIGtable}   Benchmark reactions for the evaluation of 
ILC detectors}
\begin{tabular}{|l||l|l|l|l|l|r|}
\hline
                & Process  and      & Energy      
& Observables & Target    & Detector & Notes \\
                & Final states      & (TeV)       &             
& Accuracy  & Challenge  &     \\
\hline \hline 
                &                   &             &             
&           &            &     \\
{\it Higgs}     & $ee \to Z^0h^0 \to \ell^+ \ell^- X$ & 0.35 & 
                  $\mathrm{M_{recoil}}$,\  $\sigma_{Zh}$,\  ${\rm BR}_{bb}$ &  
                  $\delta \sigma_{Zh}$ = 2.5\%, 
               $\delta {\rm BR}_{bb}$ = 1\% &  T     & \{1\}    \\
                & $ee \to Z^0h^0$, $h^0 \to b \bar b/c \bar c/\tau \tau$ 
& 0.35 & 
                  Jet flavour , jet $(E,\vec{p})$  
& $\delta {\rm M}_h$=40 MeV, \ 
                  $\delta (\sigma_{Zh} \times \mathrm{BR})$=1\%/7\%/5\% & V 
     & \{2\}\\
                & $ee \to Z^0h^0$,$h^0 \to W W^*$ & 
0.35 & ${\rm M}_Z$, ${\rm M}_W$, 
                  $\sigma_{qqWW^*}$ & 
$\delta(\sigma_{Zh} \times {\rm BR}_{WW^*})$=5\% &  C & \{3\}  \\
                & $ee \to Z^0h^0$/$h^0 \nu \bar \nu$, $h^0 \to 
\gamma \gamma$ & 1.0 & $M_{\gamma \gamma}$ & 
                  $\delta(\sigma_{Zh} \times
 {\rm BR}_{\gamma \gamma})$=5\% &  C     & \{4\}  \\
                & $ee \to Z^0h^0$/$h^0 \nu \bar \nu$, $h^0 \to 
\mu^+\mu^-$ & 1.0 & $M_{\mu \mu}$ & 
                  $5 \sigma$ Evidence for $M_h$ = 120~GeV &  T     & \{5\}  \\
                & $ee \to Z^0h^0$,$h^0 \to 
{\mathrm invisible}$ & 0.35 & $\sigma_{qq E\hspace*{-2mm}}$ & 
                  $5\sigma$ Evidence for
 $\mathrm{BR_{invisible}}$=2.5\%  &   C     & \{6\} \\
                & $ee \to h^0 \nu \bar \nu$ & 0.5 &
 $\sigma_{bb\nu\nu}$, $M_{bb}$ & 
                  $\delta (\sigma_{\nu\nu h} \times
 {\rm BR}_{bb})$ = 1\% &   C    & \{7\}   \\
                & $ee \to t \bar t h^0$ & 1.0 &
 $\sigma_{tth}$ & $\delta g_{tth}$=5\% & C & \{8\} \\
                & $ee \to Z^0h^0h^0$, $h^0 h^0 \nu \bar \nu$ & 0.5/1.0  & 
                  $\sigma_{Zhh}$, $\sigma_{\nu\nu hh}$,
 $M_{hh}$ & $\delta g_{hhh}$=20/10\% & C & \{9\} \\  \hline
{\it SSB}       & $ee \to W^+W^-$ & 0.5  & &
 $\Delta\kappa_{\gamma}, \lambda_{\gamma}$ = $2 \cdot 10^{-4}$ 
                &  V     & \{10\} \\
                & $ee \to W^+W^- \nu \bar \nu/Z^0Z^0
 \nu \bar \nu $ & 1.0 & $\sigma$ & 
                  $\Lambda_{*4},\Lambda_{*5}$ = 3~TeV &
 C & \{11\}  \\ \hline \hline
{\it SUSY}      & $ee \to \tilde{e}^+_R \tilde{e}^-_R$
 (Point 1) & 0.5 & $E_e$ & 
                  $\delta M_{\tilde \chi^0_1}$=50~MeV &  T & \{12\}\\ 
                & $ee \to \tilde{\tau}^+_1 \tilde{\tau}^-_1$, 
$\tilde{\chi}_1^+ \tilde{\chi}_1^-$ (Point 1) & 
                 0.5 & $E_{\pi}$, $E_{2 \pi}$, $E_{3 \pi}$ & 
                 $\delta (M_{\tilde \tau_1} - M_{\tilde \chi^0_1})$=200 MeV 
& T & \{13\} \\
                & $ee \to \tilde{t}_1 \tilde{t}_1$ (Point 1)& 1.0& 
& $\delta M_{\tilde t_1}$=2 GeV & & \{14\} \\
\hline
-{\it CDM}      &  $ee \to \tilde{\tau}_1^+ \tilde{\tau}_1^-$,
 $\tilde{\chi}_1^+ \tilde{\chi}_1^-$ (Point 3)  
                & 0.5 &  &  $\delta M_{\tilde \tau_1}$=1~GeV,
 $\delta M_{\tilde \chi^0_1}$=500~MeV, & F & \{15\} \\
                & $ee \to \tilde{\chi}_2^0 \tilde{\chi}_3^0$,
 $\tilde{\chi_1^+} \tilde{\chi_1^-}$ (Point 2) & 
                  0.5 & $M_{jj}$ in $jj E\hspace*{-2mm}/$,
 $M_{\ell \ell}$ in $jj \ell \ell E\hspace*{-2mm}/$ & 
                  $\delta \sigma_{\tilde \chi_2 \tilde \chi_3}$ = 4\%, 
                  $\delta (M_{\tilde \chi^0_2}-M_{\tilde \chi^0_1})$=
500~MeV &   C  & \{16\} \\
                & $ee \to \tilde{\chi_1^+} 
\tilde{\chi_1^-}$/$\tilde{\chi}_i^0 \tilde{\chi}_j^0$ (Point 5) &
                  0.5/1.0 & $ZZ E\hspace*{-2mm}/$, $WW E\hspace*{-2mm}/$ 
                & $\delta \sigma_{\tilde{\chi} \tilde{\chi}}$=10\%,
$\delta (M_{\tilde \chi^0_3} - M_{\tilde \chi^0_1})=$2~GeV & C & \{17\} \\
                & $ee \to H^0 A^0 \to b \bar b b \bar b$ 
         (Point 4) & 1.0  & Mass constrained $M_{bb}$ & 
                  $\delta M_A$=1 GeV &   C    & \{18\}   \\ \hline  
-{\it alternative} & $ee \to \tilde{\tau}_1^+ \tilde{\tau}_1^-$ 
(Point 6) & 0.5 & Heavy stable particle & 
                    $\delta M_{\tilde \tau_1}$ &  T     & \{19\} \\
{\it SUSY}  & ${\tilde \chi}^0_1 \to \gamma + E\hspace*{-2mm}/$ (Point 7) 
 & 0.5 & Non-pointing $\gamma$ & 
                 $\delta c \tau$=10\%   &   C    & \{20\} \\
{\it breaking}   & $\tilde{\chi}^\pm_1 \to \tilde{\chi}^0_1 + 
\pi^\pm_{soft}$ (Point 8) & 0.5 & Soft $\pi^\pm$ 
                above $\gamma\gamma$ bkgd & $5\sigma$ Evidence for 
$\Delta{\tilde{m}}$=0.2-2~GeV & F & \{21\} \\
\hline \hline
{\it Precision SM}   & $ee \to t \bar{t} \to 6~jets$ & 1.0  & & 
$5\sigma$ Sensitivity for 
                  $(g-2)_t/2 \leq 10^{-3}$ & V  & \{22\} \\
                 & $ee \to f \bar{f} \; (f=e,\mu,\tau;b,c)$ & 1.0 &
 $\sigma_{f \bar{f}}$, $A_{FB}$, $A_{LR}$ & 
                 $5\sigma$ Sensitivity to $M_{Z_{LR}} =$ 7~TeV & V   
   & \{23\} \\ 
{\it New Physics}    & $ee \to \gamma G$ (ADD) & 1.0 &
 $\sigma(\gamma + E\hspace*{-2mm}/)$ & $5\sigma$ Sensitivity 
& C  & \{24\} \\
                 & $ee \to KK \to f \bar{f}$ (RS) & 1.0   &   &    &   T 
   & \{25\} \\
\hline
{\it Energy/Lumi}      & $ee \to ee_{fwd}$  & 0.3/1.0   &  
    & $\delta M_{top}$=50~MeV    &   T    & \{26\} \\
{\it Meas.}     & $ee \to Z^0 \gamma$   & 0.5/1.0   &      &     &  T 
    & \{27\} \\
\hline     
\end{tabular}
\end{center}
\end{table}
\end{turnpage}

\section{DETECTOR SUBSYSTEMS}

In this section, we review the various subsystems whose performance 
contributes to the overall detector design.  We explain which 
benchmark processes are especially important to the optimization of 
each subsystem.

\subsection{Vertexing}

The branching fractions of a light Higgs boson to 
$b \bar b,\ c \bar c$ still form
the core of the case for excellent vertex detector 
performance at the ILC.  Accurate 
measurements of these branching fractions are needed
to test the Higgs mechanism of mass generation.
Recent progress in the determination of heavy quark masses 
using spectral distribution moments in semileptonic $b$ decays has provided a 
significant reduction in $M_q$ uncertainties, thus motivating 
an even better experimental accuracy.

Vertexing may play a role in $\tau$ identification in the measurement 
of the branching fraction for 
$h\to \tau^+\tau^-$ or in isolating the stau signal in 
$e^+e^-\to \tilde{\tau}_1 \tilde{\tau}_1\to \tilde \chi^0_1 
\tilde \chi^0_1 \tau^+ \tau^-$.
Precise vertexing is also useful in efficiently selecting the sign 
of a $b$ or $c$ quark through
the  measurement of the net charge of the displaced vertex.   
The vertex charge technique can resolve combinatoric ambiguities in 
the reconstruction of  $e^+e^-
\to W^+W^-$ and $e^+e^-\to t\bar t$.  It will also thus 
extend the energy scale of the new physics that can be probed 
through anomalous coupling measurements as well as 
through the measurement of the 
forward-backward left-right asymmetries in  $e^+e^- 
\to b\bar b,\ c\bar c$.

\subsection{Tracking Momentum Resolution}

The tracker momentum resolution is tested by the 
requirements of the recoil mass analysis
of the Higgs-strahlung process $e^+e^- \to Zh \to \ell^+ \ell^- X$. 
This analysis is used to measure the cross 
section for $e^+e^- \to Zh$ and the branching fraction 
of the dominant Higgs decay mode.  It also provides the best Higgs mass measurement 
when there is a large branching ratio for inclusive Higgs decays to invisible particles.
The possibility of measuring the suppressed 
$h^0 \to \mu^+ \mu^-$ decay also sets an interesting
benchmark on the tracking momentum resolution.

The measurement error for the slepton and lightest 
neutralino mass via the energy endpoint spectrum in 
$e^+e^-\to \tilde \ell^+ \tilde \ell^- \to \tilde 
\chi^0_1 \tilde \chi^0_1 \ell^+  \ell^- $ is also 
sensitive to the tracker momentum resolution when 
the decay lepton energy is greater than 100 GeV.

\subsection{Tracking Pattern Recognition}

Detection of long-lived neutral particles offers specific 
challenges to track pattern recognition.
In $e^+e^- \to ZZ$, $WW$, $hZ$ and $HA$, between 50\% and 65\% of the 
events contain at least one $K^0_S$ or $\Lambda^0$. It is not uncommon for 
these particles to carry a
significant amount of the jet energy. The jet energy resolution 
will thus be compromised if 
the charged tracks from $K^0_S$ or $\Lambda^0$ decay are missed or 
misinterpreted in the tracker pattern recognition.

Several key reactions
 have large jet multiplicities.  More generally, quarks in the final
state often lead to 
  low momentum spiralling 
particles and to highly collimated jets with 
large local track densitites. These features test 
pattern recognition and track element linking in dense environments.

\subsection{Forward Tracking}

Multiple scattering significantly degrades the 
tracker momentum resolution in the forward region. 
Analyses of forward processes such as $e^+e^-\to e^+e^-,\  
\mu^+\mu^-,\ Z\gamma,\ \nu_e \bar \nu_e h$ 
will therefore be particularly sensitive to the amount of 
material in the forward part of the tracker volume.    
The observation of the suppressed $h\to \mu\mu$ decay in 
$e^+e^-\to \nu_e \bar \nu_e h$ at $\sqrt{s}=1$~TeV 
is an excellent example of a measurement requiring good 
forward tracking momentum resolution,  since the 
$M_{\mu\mu}$ signal must be kept as narrow as possible to 
be recognised above the $\nu_e \bar \nu_e Z^*$ background.

Beamline instrumentation will provide estimates of the energy, 
luminosity,  and polarization of the electron and positron 
beams  before and/or after collision, but the quantities of 
most relevance to physics analyses are the
luminosity-weighted beam energy, luminosity and polarization.  
These luminosity-weighted quantities  can only be obtained 
directly through measurements of forward processes such as 
$e^+e^-\to e^+e^-, Z\gamma$.  Efficient 
forward tracking is especially important to reconstruct  the luminosity 
spectrum from Bhabha scattering. 
Good forward momentum resolution is needed for both reactions.

Examples of physics processes requiring excellent 
beam energy and/or luminosity measurments include 
the threshold production of the top quark  and 
the production of a  Kaluza-Klein resonance 
$e^+e^-\to KK \to f \bar f$.
As LEP was able to do for  the mass 
and width of the $Z$ boson at LEP,  the ILC will be able to make a 
precision measurement of the 
mass and width of a KK resonance, but only if a very accurate
beam energy measurement is available.

\subsection{Low Momentum Leptons}

   The region of SUSY parameter space in which the dark matter
relic density is determined by slepton co-annihilation motivates
models with a near-degeneracy of sleptons and neutralinos. 
These models give very soft final state leptons from  $e^+e^-\to \tilde \ell^+ 
\tilde \ell^- \to \tilde \chi^0_1 \tilde \chi^0_1 \ell^+  \ell^-$.
Such reactions challenge the tracker's ability to efficiently
 reconstruct charge tracks with very low
$p_T$ and to identify very low momentum leptons. In some cases, the 
lepton momenta may be below
the kinematic cut-off for the particle to reach the calorimeters 
or the muon detectors. 

\subsection{Calorimeter}     

At LEP2,  hadronically decaying $W$ and 
$Z$ bosons were reconstructed  using 
kinematic fits to improve the two-jet mass resolution. 
Kinematic constraints were also used to 
improve the mass resolution of candidate $Z$ and 
$h$ bosons in searches for $e^+e^- \to Zh$. At the ILC, kinematic
fits will continue to be useful for 4-fermion 
processes such as $e^+e^-\to W^+W^-,\ ZZ,\ 
Zh\to q\bar qq\bar q,\ l^+l^-q\bar q$, 
but they will have limited utility in 6 and 
8-fermion processes such as $e^+e^-\to Zhh,\ t\bar t,\ t\bar t h$,
and will fail altogether for processes with final state neutrinos such as 
$e^+e^-\to \nu_e \bar\nu_e h,\  \nu_e \bar\nu_e W^+W^-,\  
\nu_e \bar\nu_e ZZ$ .  

The need to distinguish hadronically decaying $W$, $Z$,  and $h$ 
bosons from one another without the benefit of kinematic 
constraints drives the present ILC jet energy resolution 
specification of $\Delta E_{jet}=30\%/\sqrt{E_{jet}}$.  
Such a resolution cannot be obtained with the calorimeter alone.
With the particle flow algorithm approach adopted by ILC detector groups, 
the jet energy is calculated by combining the sum of the 
tracker momenta of charged particles with the
sum of the energies of showers not associated with charged particles. 
In this approach,   
a premium is placed on separating charged particle showers
showers initiated by photons and by neutral hadrons.  This leads to 
strong requirements on the calorimeter parameters
such as the inner radius, the radiator material, 
the pixel size, and the number of layers.

\subsection{$W$,$Z$, $h$ Boson Separation}

In the study of the strong symmetry breaking processes 
$e^+e^-\to 
\nu_e \bar\nu_e W^+W^-,\  \nu_e \bar\nu_e ZZ$ and in SUSY
 processes such as 
$e^+e^-\to \tilde{\chi}_1^+ \tilde{\chi}_1^-\to 
\tilde{\chi}_1^0 \tilde{\chi}_1^0 W^+W^-$ and
$e^+e^-\to \tilde{\chi}_2^0 \tilde{\chi}_2^0
\to \tilde{\chi}_1^0 \tilde{\chi}_1^0 ZZ$, it is 
important to  distinguish the $W^+W^-$ final state from $ZZ$.  
In the $WW$ fusion production of Higgs, 
$e^+e^-\to \nu_e \bar\nu_e h$, the Higgs signal must be 
separated from a 
background of $e^+e^-\to e^- \bar\nu_e W^+,\  e^+e^-Z,\  
\nu_e \bar\nu_e Z$ by the direct reconstruction of 
the Higgs mass.  The signal for invisible decays of the 
Higgs,  $e^+e^-\to  Zh\to q\bar q +E\hspace*{-2mm}/$,  
is obtained by requiring that the reconstructed $q\bar q$
mass be close to the $Z$ mass. and that the recoil mass obtained
from the  $q\bar q$  2-jet system be close to the Higgs mass.
When measuring the Higgs boson self-coupling,  the signal
 for the $WW$ fusion production of two Higgs bosons, 
$e^+e^-\to \nu_e \bar\nu_e hh$, must be separated from a background of
$e^+e^-\to \nu_e \bar\nu_e W^+W^-,\  \nu_e \bar\nu_e ZZ,\ 
e^- \bar\nu_e ZW^+$, etc.
In all of these cases the jet energy resolution plays a 
central role.

\begin{table} 
\begin{center}
\caption{\label{xxtable}  Table of relations between the benchmark 
physics processes and parameters of detector subsystems}

\begin{tabular}{|l|c|c|c|c|c|c|c|c|c|c|c|c|c|c|}
\hline
       Process  &  {\bf V}ertex & \multicolumn{2}{|c|}{{\bf T}racking} &
 \multicolumn{2}{|c|}{{\bf C}alorimetry} 
                & \multicolumn{2}{|c|}{{\bf F}wd} & {\bf Very} Fwd &
 \multicolumn{5}{|c|}{{\bf I}ntegration} & {\bf P}ol. \\ \hline
                &  $\sigma_{IP}$ &  $\delta p/p^2$ & $\epsilon$ &
 $\delta E$ & $\delta \theta$, $\delta \phi$ &
Trk & Cal & $\theta^e_{min}$ & $\delta E_{jet}$ & $M_{jj}$ & 
$\ell$-Id & $V^0$-Id & $Q_{jet/vtx}$ & \\
\hline \hline 
$ee \to Zh \to \ell \ell X$ & & x & &  &  & & & & & & x & & & \\
$ee \to Zh \to jjbb$ & x & x & x & & & x & & & & x & x & & & \\
$ee \to Zh$,$h \to bb/cc/\tau \tau$ & x &  & x & & & & & & & x & x & & & \\
$ee \to Zh$,$h \to WW$ & x & & x & & x & & & & x & x & x & & & \\
$ee \to Zh$, $h \to \mu \mu$ & x & x & & & & & & & & & x & & & \\ 
$ee \to Zh$, $h \to \gamma \gamma$ & & & & x & x & & x & & & & & & & \\ 
$ee \to Zh$,$h \to {\mathrm invisible}$ & & & x & & & x & x & & & & & & & \\
$ee \to \nu\nu h$ & x & x & x & x & & & x & & & x & x & & & \\
$ee \to tth$       & x & x & x & x & x & & x & x & x & & x & & & \\
$ee \to Zhh$, $\nu\nu hh$ & x & x & x & x & x & x & x &  & x & x & x &
 x & x & x \\ \hline \hline
$ee \to WW$        & & & & & & & & & & x & & & x & \\
$ee \to \nu\nu WW/ZZ$ & & & & & & x & x & & x & x & x & & & \\
\hline \hline
$ee \to \tilde{e}_R \tilde{e}_R$ (Point 1) & & x & & & & & & x & & & x & & & x \\
$ee \to \tilde{\tau}_1 \tilde{\tau}_1$ & x & x & & & & & & x & & & & & & \\
$ee \to \tilde{t}_1 \tilde{t}_1$ & x & x & & & & & &  & x & x & & x & & \\
\hline
$ee \to \tilde{\tau}_1 \tilde{\tau}_1$ (Point 3)  & x & x & & & x & x & x & x &
 x & & & & & \\
$ee \to \tilde{\chi}_2^0 \tilde{\chi}_3^0$ (Point 5) & & & & & & & & & x & x & & & & \\
$ee \to H A \to bbbb$      & x & x & & & & & &  & & x & x & & & \\
\hline  
$ee \to \tilde{\tau}_1 \tilde{\tau}_1$ & & & x & & & & & & & & & & & \\
$\tilde{\chi}^0_1 \to \gamma + E\hspace*{-2mm}/$ & & & & & x & & & & & & & & 
& \\
$\tilde{\chi}^\pm_1 \to \tilde{\chi}^0_1 + \pi^\pm_{soft}$ & & & x & & & &
 & x & & & & & & \\
\hline \hline
$ee \to tt \to 6~jets$    & x & & x & &  & & & & x & x & x & & & \\
$ee \to ff \; [e,\mu,\tau;b,c]$ & x & & x & & & & x & & x & & x & & x & x \\
$ee \to \gamma G$ (ADD) & & & & x & x & & & x &  & & &  & & x \\  
$ee \to KK \to f \bar f$ & & x & & & & &  & &  & & x & &  &  \\
\hline
$ee \to ee_{fwd}$    & & & & & & x & x & x & & & & & & \\
$ee \to Z\gamma$     & & x & & x & x & x & x & & & & & & & \\
\hline  
   
\end{tabular}
\end{center}
\end{table}

\subsection{Multi-Jet Combinatorics}

Good calorimetry  can help reduce the combinatorics in 
reactions with four or more jets.  Examples for which this
is particularly important are the measurements of the 
Higgs branching fraction to $WW^*$
using
$e^+e^-\to Zh\to Z W W^*\to q\bar q q \bar q l \nu$,  the 
measurement of the 
Higgs self-coupling 
using $e^+e^-\to Zhh\to q\bar q b \bar b b \bar b$, 
and the measurement of  the top quark Yukawa coupling using 
$e^+e^-\to t\bar t h\to b\bar b b\bar b q\bar q q\bar q$.
In all of these cases, 
the likelihood that jets will be correctly assigned to 
their parent particles will depend on how well the 
calorimeter reconstructs $W$ $Z$, and $h$ boson  masses.

\subsection{Heavy Particle Mass Measurements}

Although kinematic constraints can be used to significantly 
improve the Higgs mass resolution in the processes $e^+e^-\to 
 Zh\to q\bar q b\bar b$ and
$e^+e^-\to  HA\to b\bar b b\bar b$, the calorimeter performance
 may still play an important role in determining the ultimate 
Higgs mass precision.

\subsection{Photons}

In the strategy of particle flow calorimeter, the electromagnetic
calorimeter should be  designed to optimize the separation of charged showers 
from neutral showers.  However, it is important in doing this
not to overlook traditional calorimeter figures of merit such 
as the intrinsic resolution for electromagetic showers.    
An example of a measurement
requiring very good electromagnetic calorimeter energy resolution 
is the measurement of the Higgs branching ratio to two photons.  
Initial measurements of this
branching ratio can be performed at $\sqrt{s}=350$~GeV using the 
Higgs-strahlung process;  later the error on the branching fraction can be 
improved using $e^+e^-\to \nu_e \bar\nu_e h$
at $\sqrt{s}=1$~TeV.     

The ability to efficiently detect soft photons over most of  the 
solid angle is important for the measurement of graviton production 
in association  with a photon, $e^+e^-\to \gamma G$, seen as 
$\gamma + E\hspace*{-2mm}$.

The fine granularity of an ILC calorimeter can help identify photons 
that do not point back to the interaction point.  For example, in 
GMSB models of SUSY the neutralino may decay inside the detector via 
$\tilde{\chi}^0_1 \to \gamma + E\hspace*{-2mm}/$, producing two photons 
appearing in the detector superimposed on each SUSY event.

\subsection{Far Forward Detector}

The far forward detector should be able to veto electrons down 
to an angle of a few mrads in the presence of a large $e^+e^-$ 
pair background.  This capability is needed in general 
to suppress background from $\gamma\gamma\to f\bar f$ in any 
analysis with missing energy in the final state. It takes 
on added importance given the possiblity that some SUSY particles
 may be nearly degenerate with the LSP.  Examples
of processes requiring superlative rejection of  the 
$\gamma\gamma\to f\bar f$ background include
$e^+e^-\to \tilde\tau_1 \tilde\tau_1\to \tilde{\chi}_1^0 
\tilde{\chi}_1^0 \tau^+\tau^-$ and 
$e^+e^-\to \tilde{\chi}_1^+ \tilde{\chi}_1^-\to \tilde{\chi}_1^0 
\tilde{\chi}_1^0 \pi^+ \pi^-$ where
the $\tilde\tau_1$ or  $\tilde{\chi}_1^+$ is nearly degenerate 
with the $\tilde{\chi}_1^0$.

\section{SELECTING A REDUCED BENCHMARK LIST} 

The broad set of physics reactions identified
above provides detailed coverage of the
detector performance requirements for an ILC detector.
However, the list of reactions is rather long.  Thus, it is useful
to call out a smaller subset of benchmarks which emphasize
key aspects of the detector performances---vertexing, tracking,
calorimetry, very forward instrumentation and
integration. There are certainly several such subsets
which could be chosen. In our choice, we give special weight to
those reactions whose analysis maintains a simple
relation between the physics measurement and the
basic detector parameters.  We also give weight to those
reactions that are key to the ILC physics case and
emphasise the extended capabilities that the ILC will provide
with respect to the LHC.

The Higgs sector offers several
reactions. The process $e^+e^- \to Zh \to \ell \ell X$ tests
momentum resolution for energetic, isolated charged particle
tracks. It presents a challenging accuracy target for
$\sigma_{hZ}$, which is justified because this 
represents the common normalisation
for the extraction of Higgs couplings. The Higgs branching
fraction measurements  $e^+e^- \to Zh$, $h \to cc$, $ \tau \tau$,
$WW^*$ probe the detector capability in tagging heavy
flavours, which
drives the vertex performance. In particular, it gives stringent tests
of charm di-jet tagging
under a dominant $b\bar b$ background, and of $\tau$ tagging by single
particle impact parameter measurement.
The reaction $e^+e^- \to Zhh$,
 which give access to the Higgs self-coupling and
the reconstruction of the Higgs potential, tests energy
 flow and di-jet mass resolution to discriminate the
$Zhh$ signal from the $ZZZ$ background. It also tests $b$ tagging
efficiency in complex six-jet final states, and it provides an
example in which vertex charge can be used to resolve combinatoric 
ambiguities in jet assignment.  

Among the variety of  SUSY reactions, $e^+ e^- \to \tilde e_R \tilde e_R $
gives a  sharp test of  lepton tagging and momentum resolution.  The reaction
$e^+ e^- \to \tilde \tau_1 \tilde \tau_1$
in the co-annihilation region of cosmologically
interesting supersymmetric parameters offers severe challenges to
charged particle detection, due to the small splitting between the 
$\tilde \tau_1$ and the $\tilde{\chi}_1^0$,  and to the design of the far
forward region, due to the need to detect forward electrons to suppress
the $e^+e^- \to e^+e^-\tau^+\tau^-$ two-photon background.
Reactions containing the decays 
$\tilde{\chi}_1^\pm \to W^\pm \tilde{\chi}_1^0$ and 
$\tilde{\chi}_2^0 \to Z^0 \tilde{\chi}_1^0$ test the energy flow performance 
through on-shell $W$ and $Z$ identification and reconstruction.

Two-fermion final states at the highest energy provide an
excellent test of tracking and calorimetric
performances with high local particle density. The study
of $e^+e^- \to f \bar f$ may offer a valuable
window on physics well beyond the $\sqrt{s}$ limit by
electroweak fits to precision observables. This
study requires good $b$ and $c$ jet flavor tagging, in an 
almost democratic flavor mixture, and measurements of jet and vertex charge
 for computing forward-backward asymmetries.  Polarization-dependence is
large, so the analysis also benefits from the use and precision measurement
of beam polarization.

Finally, samples of single particles, produced over the whole polar angle 
and a wide energy range, and di-jet events will complement these benchmarks 
and highlight the deterioration in efficiency and resolution for particles in
jets and complex events. These samples will enable tests of
reconstruction in a simple environment. They will  provide a basis for
defining the detector response in terms of efficiency
and resolution and will produce well-motivated parametrizations to be
used in physics studies carried out with fast simulation.

In this way, we have arrived at the following proposal for a reduced
list of benchmark reactions for initial study.  Our list includes
reactions that test all elements from the menu of basic detector
capabilities.

\begin{enumerate}
\setcounter{enumi}{-1}

\item Single $e^{\pm}$, $\mu^{\pm}$, $\pi^{\pm}$, $\pi^0$,
$K^{\pm}$, $K^0_S$, $\gamma$,
$0 < | \cos \theta| <1$, $0 < p < 500$~GeV

\item $e^+e^- \to f \bar f$, $f = e$, $\tau$, $u$, $s$, 
$c$, $b$ at $\sqrt{s}$=0.091, 0.35, 0.5 and 1.0~TeV;

\item $e^+e^- \to Z^0h^0 \to \ell^+ \ell^- X$,
$M_h$ = 120~GeV at $\sqrt{s}$=0.35~TeV;

\item $e^+e^- \to Z^0h^0$, $h^0 \to c \bar{c}$, $ \tau^+ \tau^-$,
$WW^*$, $M_h$ = 120~GeV at $\sqrt{s}$=0.35~TeV;

\item $e^+e^- \to Z^0h^0h^0$, $M_h$ = 120~GeV at $\sqrt{s}$=0.5~TeV;

\item $e^+ e^- \to \tilde e_R^+ \tilde e_R^-$ at Point 1 at
$\sqrt{s}$=0.5~TeV;

\item $e^+ e^- \to \tilde \tau_1^+ \tilde \tau_1^-$,
at Point 3 at $\sqrt{s}$=0.5~TeV;

\item $e^+ e^- \to \tilde{\chi}_1^+ \tilde{\chi}_1^-$/$\tilde{\chi}^0_2 
\tilde{\chi}^0_2$ at Point 5 at $\sqrt{s}$=0.5~TeV;

\end{enumerate}

This list of reactions seems still to be rather lengthy, especially since
each item contains a number of distinct physics measurements.  However,
we believe that, for the initial stages of detector design and
optimization, this set of physics topics can be surveyed at a useful
level. In preparing this list, we have been careful to
include only reactions that
have already been studied extensively and for which
established analysis algorithms exist.  These algorithms, based on
reconstructed physics objects such as electrons, muons, charged hadrons,
photons, and neutral hadrons, are to a very good approximation
independent of the details of the detector design.  In fact, they could be
standardized and shared among the detector concept groups.
Once a concept group has developed the software needed to convert fully
simulated raw data to reconstructed electrons, muons, charged hadrons, photons,
and neutral hadrons, it can utilize the common standardized physics algorithms
to quickly survey the physics capability of its design.
Analyses at this level, while not achieving the ultimate optimized
performance for each detector, would come close enough to guide the
detector design process and also to realistically predict the physics capabilities
of the ILC experiments.

\section{CONCLUSIONS}

In this note, we have reviewed benchmark physics processes for the 
ILC that give broad coverage of the requirements that ILC detectors must
satisfy.  We have described the mapping between these reactions and the 
capabilties they require in the various detector subsystems.  Finally, 
we have provided guidelines for the selection of smaller sets of benchmarks
for the early stages of detector development, and we have suggested an 
especially economical subset on which to begin the detector performance 
studies.

\begin{acknowledgments}
This paper has been prepared on request of the ILC Worldwide Study. 
The authors wish to thank Hitoshi Yamamoto for extensive discussion and 
Yasuhiro Sugimoto and Graham Wilson for their contributions to the LCWS05 
benchmark parallel session. The work of M.B. was supported by the Director, 
Office of Science, Office of Basic Energy Sciences, of the U.S. Department 
of Energy under Contract No. DE-AC02-05CH11231. The work of T.B. and M.P. 
has been supported by the US Department of Energy, contract DE-AC02-76SF00515. 
The work of Y.O. has been supported by a Grant-in-Aid of the 
Ministry of Education, Culture, Sports, Science, and Technology, Government of
Japan, Nos.~13640309, 13135225, 16081211, and 17540286.
\end{acknowledgments}

\newpage

\noindent
{Notes to Table~II}

{\small

\noindent
\{1\}  The target is to determine the Higgs-strahlung cross section 
with 2.5\% accuracy, in addition the 
Higgs mass should be obtained to $\delta {\rm M}_h$=100 MeV.

\noindent
\{2\}  The Higgs branching fractions can be extracted by a 
fit to the jet flavour tagging response on 
candidate $h^0Z^0$ events.  The target accuracies correspond 
to the estimated theoretical accuracies assuming the $b$ and 
$c$ quark masses to be determined at $B$ factories to $\pm$ 50~MeV 
and $\pm$ 30~MeV, respectively~\cite{Aubert:2004aw,Buchmuller:2005zv}
or alternatively from 
heavy quark production in low-energy $e^+e^-$ 
annihilation~\cite{Kuhn:2001dm}.

\noindent
\{3\} The target precision on BR($h^0 \to W^+W^-$) is set by the 
requirement of comparable accuracies in 
$\Gamma_{h \to W}$ (measured in $WW$ fusion) and in the branching 
fraction, for the extraction of 
the Higgs total decay width~\cite{Battaglia:2000jb}.

\noindent
\{4\}  The analysis should be performed at $\sqrt{s}$=1~TeV, in the sum 
of the $\gamma \gamma j j$ and $\gamma \gamma \nu \nu$, sensitive to both 
Higgs-strahlung and fusion production processes, as 
in~\cite{Boos:2000bz,Barklow:2003hz}.
The target precision is set by requiring the Higgs total width to be determined, 
using the combination of  $h \to \gamma \gamma$ and $\gamma \gamma \to h$ at a 
photon collider, to an accuracy comparable to that 
using the $WW$ coupling (see note \{3\}).

\noindent
\{5\} Observe a $> 5 \sigma$ signal of $h \to \mu \mu$ assuming SM couplings
 for a $M_h$ = 120~GeV.

\noindent
\{6\} The target is to obtain a 5~$\sigma$ signal for $h \to 
{\mathrm{invisible}}$ for 
an invisible branching fraction of 2.5\%, from a $j j E\hspace*{-2mm}/$ 
analysis~\cite{Schumacher:2003ss}.

\noindent
\{7\}  A challenge for $b$-tagging and $M_{jj}$ in a case where global
kinematic fitting is not possible.

\noindent
\{8\} The target is to measure the top Yukawa coupling $g_{tth}$ to a 
statistical accuracy of 5\% from the analysis of $e^+e^- \to t \bar t h$ at 
$\sqrt{s}$=1~TeV.  

\noindent
\{9\}  Improved techniques in double Higgs-strahlung and WW fusion and 
availability of polarised 
     beams are expected to reduce the individual errors for the triple
 Higgs
     coupling to about 15\% in the two channels so that the combined
     accuracy should finally reach the 10\% 
limit~\cite{Battaglia:2001nn,Yasui:2002se,yamashita:2004}. 

\noindent
\{10\}  The analysis is carried out for exact SU(2)$\!\times\!$U(1)
 invariance. The 
     five CP-invariant static electroweak parameters of the $W^\pm$ bosons,
 {\it i.e.} 
     the Z-charge, the two magnetic dipole and electric quadrupole moments 
can be 
     expressed in terms of just three parameters $\Delta g_Z$, 
$\Delta\kappa_Z$ and 
      $\lambda_Z$ as in Ref.~\cite{kuroda}.

\noindent
\{11\}  Interactions among $W$ and $Z$ bosons in scenarios of strong
     electroweak symmetry breaking are generally described by chiral
     expansions. The expansion parameters, $\{v/\Lambda_{*}\}^2$, are
     expressed by strong interaction scales which may extend up to values
     of $4 \pi v =$ 3 TeV. Beyond $WW$ energies of 3~TeV resonances should
     be formed. In theories respecting SU(2) isospin invariance two
     parameters determine the amplitudes for [quasi-]elastic
     $WW \to WW, ZZ$ scattering, $\Lambda_{*4}$ and $\Lambda_{*5}$,
     see Ref.~\cite{Boos}. 
     A 1~TeV ILC ought to cover the entire  $\Lambda_{*4}$ 
     and $\Lambda_{*5}$ parameter region below 3~TeV. The reach refers to an 
     analysis based on optimal observables~\cite{Moenig}.

\noindent
\{12\} SUSY parameters of Point~1 of Table~1, corresponding to  
     update of mSUGRA SPS1a/LCC~1 point~\cite{Allanach:2002nj,lccosmo}, 
     now SPS1a'. The target is to extract the 
     lightest neutralino mass to the specified accuracy from the analysis 
     of the 
     lepton energy spectrum, assuming to know the selectron mass to 
$\pm$~50~MeV
     from a dedicated threshold scan.

\noindent
\{13\} SUSY parameters of Point~1 of Table~1, the 
$\tilde \tau_1$ to $\tilde{\chi}^0_1$ mass difference can be obtained by 
a study of the energy distribution of the hadronic system in the decays 
$\tilde \tau_1^{\pm} \to \tilde{\chi}^0_1 \tau^{\pm}$, $\tau^{\pm}
\to \pi^{\pm} \nu$, $\rho^0 \nu$, $a_1^{\pm} \nu$, which give one, two and 
three pion final states~\cite{Martyn:2004jc}. 

\noindent
\{14\} A challenge for precision kinematic endpoint measurement from
jets.

\noindent
\{15\} SUSY parameters in co-annihilation region, Point~3 of Table~1,
 corresponding to mSUGRA Point D' of~\cite{Battaglia:2003ab}. 
The target accuracies on the $\tilde \tau$ and neutralino mass are 
required to determine $\Omega_{\chi} h^2$ to a statistical 
accuracy comparable to that of CMB experiments. 
The $\tilde \tau$ mass can be obtained to a $\pm$500~MeV accuracy from a 
dedicated threshold scan. The $\tilde \tau \tilde \tau \to \tau \tilde \chi 
\tau \tilde \chi$ is contaminated by $ee \to \tau \tau ee$ which requires 
low angle $e$ tagging and, possibly, $\mu$/$\pi$ id in the 
very forward instrumentation~\cite{Bambade:2004tq}.

\noindent
\{16\} SUSY parameters in focus point region, Point~2 of Table~1, 
corresponding to  mSUGRA Point LCC~2 of~\cite{lccosmo}. 
Gauginos decays into ligther states and virtual $W$ and $Z$ bosons. 
The target accuracies on the gaugino mass differences
are required to determine $\Omega_{\chi} h^2$ to a 5-10~\% 
statistical accuracy. The mass differences can be extracted by the 
kinematical upper limit of the $M_{jj}$ and 
$M_{\ell \ell}$ in the $jj \nu \nu \tilde \chi \tilde \chi$ and 
$jj \ell \ell \tilde \chi \tilde \chi$ final states, respectively. 
The difference in the kinematics allows to distinguish 
$ee \to \tilde \chi_2 \tilde \chi_3$ from $ee \to \tilde \chi_1 \tilde \chi_3$.
The cross section $\sigma_{\tilde \chi_2 \tilde \chi_3}$ 
can be extracted, in the $jj \ell \ell$ channel~\cite{Gray:2005ci}.  

\noindent
\{17\} SUSY parameters for non universal Higgs masses (NHUM) model, 
from~\cite{DeRoeck:2005bw}, 
Point~5 of Table~1. This point has real $W$ and $Z$ bosons produced 
in heavier gaugino cascade decays. 

\noindent
\{18\} SUSY parameters in $A$ annihilation funnel, Point~4 of Table~1, 
correspondinding to mSUGRA Point LCC~4 of~\cite{lccosmo}. The target 
accuracy on $M_A$ is required to extract $\Omega_{\chi} h^2$ to 
5-10\% statistical accuracy~\cite{Battaglia:2004gk}.

\noindent
\{19\} SUSY parameters for Point~6 of Table~1, from~\cite{DeRoeck:2005bw}, 
with gravitino LSP and quasi-stable next-to-lightest supersymmetric 
particle~\cite{Feng:2004yi}.
Experimental signature is long-lived particle traversing the detector 
as heavy ionizing particle. The track curvature in the magnetic field 
can be used to determine the $\tilde \tau$ mass. The reference mass is 
$M_{\tilde{\tau}_1} =$ 150~GeV. 

\noindent
\{20\} SUSY GMSB scenario, the final state particle in SUSY decay chain
is the gravitino, albeit with a mass of less than 
keV~\cite{Ambrosanio:1999xk}. 
The benchmark point parameters are $M_{messenger}$ = 240~TeV, $\Lambda$  
= 110~TeV, $\sqrt{F}$ = 0.8~PeV, $M_{\tilde G}$    = 0.15~keV, 
$M_{\tilde \chi^0_1}$ = 150~GeV, $M_{\tilde \tau}$ = 200~GeV.

\noindent
\{21\} SUSY AMSB scenario: in this scheme, the charginos $\tilde{\chi}_1^\pm$
     are nearly mass degerate with the lightest neutralino
     $\tilde{\chi}^0_1$. Chargino pair production therefore generates two
     soft hadrons, {\it viz.} $e^+e^- \to \tilde{\chi}_1^+
     \tilde{\chi}_1^- \to \pi^+ \pi^- + E\hspace*{-2mm}/$,
     see Ref.~\cite{Gunion:2001fu}. SUSY parameters of point 
     SPS9~\cite{Allanach:2002nj}: $m_0$ = 450~GeV, $m_{3/2}$ = 60000~GeV, 
     $\tan \beta$ = 10, Sign($\mu$) = +1 with $M_{top}$ = 178~GeV, corresponding  
     to $M_{\tilde{\chi}^0_1} =$ 180.88~GeV and 
     $M_{\tilde{\chi}^\pm_1} =$ 181.05~GeV in {\tt ISAJET.7.69}. 

\noindent
\{23\} The accuracy in fermion pair production properties at ILC can be 
     quantified by studying the mass reach for the heavy neutral vector boson
     $M_{Z_{LR}}$ in left-right symmetric extensions of the Standard
     Model. The virtual effects should be determined at ILC so well
     that the mass range should be tripled compared with direct
     production at LHC~\cite{RS}. The $A_{FB}$ measurement is particularly
     sensitive to jet/vertex charge determination. 

\noindent
\{24\} Single photon signature of $ee \to \gamma G$ in ADD scenario for 
     number of extra dimensions $\delta$ = 4, $M_D$= 4~TeV, see 
     Refs.~\cite{Giudice:1998ck,Mirabelli:1998rt}
     Beam polarization 
     (${\cal{P}}(e^-) \simeq 0.8$, ${\cal{P}}(e^+) \simeq 0.6$) is needed for 
     suppressing the $ee \to \nu \nu \gamma$ background~\cite{Wilson:2001me}.

\noindent
\{25\} Two fermion final states of Kaluza Klein graviton excitation production in 
     Randall Sundrum model~\cite{Davoudiasl:1999jd,Rizzo:2004kr}. The model 
     parameters are $m_1$ = 1~TeV and $k / \bar M_{Pl}$ = 0.05, where $k$ is 
     the curvature parameter and $\bar M_{Pl}$ the reduced Planck mass. The 
     decay is implemented in {\tt Pythia}, since version 6.2, at process 
     {\tt ISUB = 391}. In version 6.3, this benchmark corresponds to 
     {\tt PMAS(PYCOMP(5000039),1) = 900.d0} and {\tt PARP(50) = 0.05d0}.   
     The aim is to determine the resonance mass and width and to test its 
     decay branching fractions to leptons and quarks with accuracies comparable
     to those obtained at LEP/SLC for the $Z^0$. 

\noindent
\{26\} Forward Bhabha scattering, partly acoplanar due to initial and
     final-state photon radiation, serves to measure energy and
     luminosity, accounting thereby for the effects of soft real
     photon radiation due to beamstrahlung \cite{DJM}. 
     Observed {\it in situ}, these measurements are a necessary
     complement to using external spectrometers for integral energy
     measurements before and after the bunch collisions. 

\noindent
\{27\} Beam energy determination by the $e^+e^-$ annihilation to $Z \gamma$.
}

\end{document}